\documentclass[sigconf]{acmart}
\usepackage{enumitem}
\usepackage{adjustbox}
\usepackage{booktabs}
\usepackage{graphicx}  
\usepackage{subcaption} 
\usepackage{enumitem}

\renewcommand\footnotetextcopyrightpermission[1]{} 

\AtBeginDocument{%
  }

\setcopyright{acmlicensed}
\copyrightyear{2018}
\acmYear{2018}
\acmDOI{XXXXXXX.XXXXXXX}
\acmConference[Conference acronym 'XX]{Make sure to enter the correct
  conference title from your rights confirmation email}{June 03--05,
  2018}{Woodstock, NY}
\acmISBN{978-1-4503-XXXX-X/2018/06}

\begin{document}

\title{Generative Chain of Behavior for User Trajectory Prediction}

\author{Chengkai Huang}
\affiliation{%
  \institution{UNSW and Macquarie University}
  \city{Sydney}
  \country{Australia}}
\email{chengkai.huang1@unsw.edu.au}

\author{Xiaodi Chen}
\affiliation{%
  \institution{University of New South Wales}
  \city{Sydney}
  \country{Australia}
}

\author{Hongtao Huang}
\affiliation{%
 \institution{University of New South Wales}
 \city{Sydney}
 \country{Australia}}

\author{Quan Z. Sheng}
\affiliation{%
  \institution{Macquarie University}
    \city{Sydney}
      \state{NSW}
  \country{Australia}}

\author{Lina Yao}
\affiliation{%
  \institution{UNSW and CSIRO’s Data61}
    \city{Sydney}
  \state{NSW}
  \country{Australia}}

\renewcommand{\shortauthors}{Trovato et al.}
\newcommand{\Model}{LPDO}

\begin{abstract}
Modeling long-term user behavior trajectories is essential for understanding evolving preferences and enabling proactive recommendations. However, most sequential recommenders focus on next-item prediction, overlooking dependencies across multiple future actions. We propose Generative Chain of Behavior (GCB), a generative framework that models user interactions as an autoregressive chain of semantic behaviors over multiple future steps. GCB first encodes items into semantic IDs via RQ-VAE with k-means refinement, forming a discrete latent space that preserves semantic proximity. On top of this space, a transformer-based autoregressive generator predicts multi-step future behaviors conditioned on user history, capturing long-horizon intent transitions and generating coherent trajectories. Experiments on benchmark datasets show that GCB consistently outperforms state-of-the-art sequential recommenders in multi-step accuracy and trajectory consistency. Beyond these gains, GCB offers a unified generative formulation for capturing user preference evolution.
\end{abstract}

\ccsdesc[500]{Information systems~Recommender systems}

\keywords{Generative Recommendation, User Trajectory Prediction}



\maketitle

\section{Introduction}

User behavior in online platforms unfolds as a trajectory of interactions rather than isolated clicks, where each action reflects evolving preferences over time \citep{CF1,CF2,foundationsurvey,huang2025pluralistic,huang2025embedding}. While sequential recommender systems have made substantial progress, most methods are still optimized for one-step next-item prediction \citep{wang2021survey,ye2025beyond,ye2025gaussianmixtureflowmatching,www23huang,sigir23huang}. This single-step view limits their ability to reason about how user intent develops over multiple future actions, which is crucial for proactive recommendation, retention modeling, and long-term planning.

Early work extended collaborative filtering with temporal and sequential structures such as personalized Markov chains and time-aware factorization \citep{FPMC}. Deep models based on recurrent networks, self-attention, and self-supervised learning further improved the modeling of historical interactions \citep{SARS}. In parallel, generative and diffusion-based recommenders have begun to treat interactions as discrete tokens and model stochastic generation of future items \citep{lin2024survey}. However, these approaches typically focus on the immediate next item and rely on dense item embeddings that can be hard to scale and may lack semantic structure in large catalogs \citep{BillionEmbedding}.

To address these limitations, we propose \emph{Generative Chain of Behavior} (GCB), a generative framework for \emph{multi-step user trajectory prediction}. GCB first encodes each item into a \emph{semantic ID}, a discrete and structured representation learned via residual quantization (RQ-VAE) followed by k-means refinement. This construction produces a compact latent vocabulary where semantically related items are mapped to nearby codes, which stabilizes the generative process over long horizons. On top of this discrete semantic space, GCB employs a transformer-based autoregressive generator \citep{Transformer} that sequentially produces future behavior tokens conditioned on a user's historical interactions. Instead of optimizing only for one-step prediction, GCB is trained to generate an entire chain of future behaviors over $k$ steps, explicitly modeling the dependencies among future actions and the evolution of user intent.

This formulation offers two main benefits. First, by operating in a discrete semantic ID space, GCB mitigates issues such as embedding collapse and enables stable multi-step generation that remains coherent with both historical context and item semantics. Second, viewing user behavior as a chain of semantic transitions allows GCB to capture temporal intent drift and multi-modal future patterns that are not well represented by single-step objectives or deterministic predictors. 
Our main contributions are threefold:
\begin{itemize}
    \item We introduce \textbf{Generative Chain of Behavior}, a generative framework that directly targets multi-step user trajectory prediction instead of only next-item recommendation.
    \item We design a semantic ID representation that combines RQ-VAE and k-means clustering to form a discrete, semantically meaningful space for autoregressive behavior generation.
    \item We conduct extensive experiments on standard benchmarks, demonstrating that GCB yields more accurate and coherent long-horizon trajectories than state-of-the-art sequential and generative recommenders.
\end{itemize}

\begin{figure}
    \centering
    \includegraphics[width=0.99\linewidth]{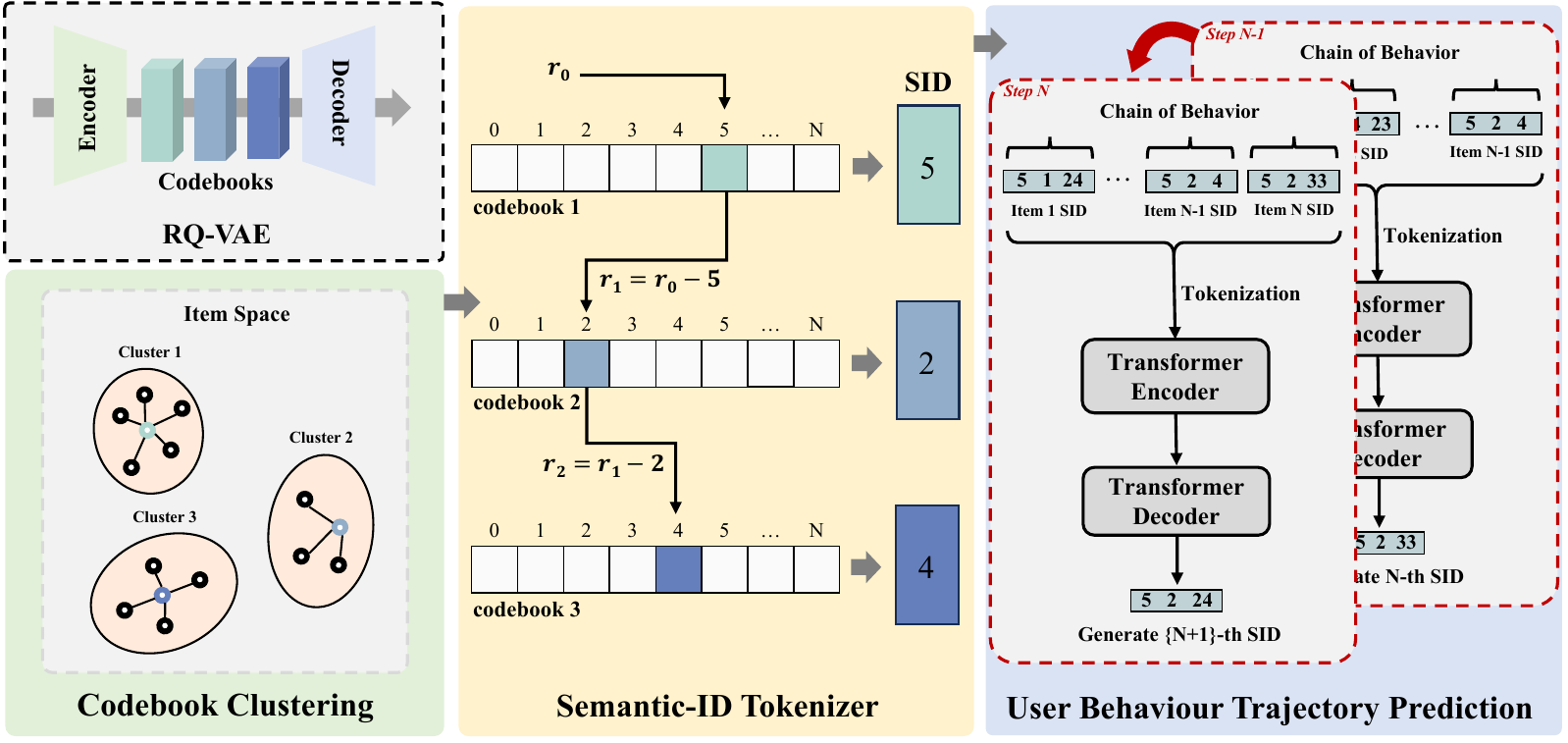}
    \vspace{-1em}
    \caption{The overview of our GCB framework.}
    \label{fig:main_fig}
\end{figure}

\section{Problem Formulation}



In this paper, we consider an item universe $\mathcal{I}=\{1,2,\dots,M\}$ of size $M$. For any user $u$, the observed interaction history up to time $t$ is represented as:
\begin{equation}
H_u = (i_{u,1}, i_{u,2}, \dots, i_{u,n}) \in \mathcal{I}^n,
\end{equation}
where $i_{u,j}$ denotes the $j$-th item interacted with by user $u$, and $n$ indicates the length of the historical sequence.

Given a user history $H_u=(i_{u,1},\dots,i_{u,n})$, the objective is to predict the following $k$ items:
\begin{equation}
S_u = (i_{u,n+1}, i_{u,n+2}, \dots, i_{u,n+k}),
\end{equation}
where for each future step $j\in\{1,\dots,k\}$, the model returns a ranked list of $K$ candidate items:
\begin{equation}
A_{u,j} = (a_{u,j,1}, a_{u,j,2}, \dots, a_{u,j,K}),
\end{equation}
such that $A_{u,j}\subseteq\mathcal{I}$ and $|A_{u,j}|=K$. The evaluation assesses whether the ground-truth item $i_{u,n+j}$ is included in the corresponding candidate set $A_{u,j}$.

\section{Methodology}

In this section, we introduce the proposed framework, Generative Chain of Behavior (GCB) for multi-step user behaviour trajectory prediction. As shown in Figure \ref{fig:main_fig}, the method consists of two core components. The first component is a Semantic-ID encoder that maps raw item identifiers into a compact and semantically structured latent space using a Residual Quantized Variational Autoencoder (RQ-VAE) trained with a k-means codebook initialization. The second component is a generative model that produces long-horizon user behaviour trajectories in a coarse-to-fine manner. This model integrates user history and temporal step information to learn a distribution over future interaction sequences.

\subsection{Semantic-ID Encoding with RQ-VAE}

\subsubsection{Residual Quantized VAE}

Following \cite{rajput2023recommender}, for each item $i \in \mathcal{I}$, we first obtain a continuous representation $x_i \in \mathbb{R}^{d_x}$ from metadata, textual features, or a randomly initialized embedding. An encoder network $f_\theta(\cdot)$ maps $x_i$ into a latent vector:
\begin{equation}
z_i = f_\theta(x_i) \in \mathbb{R}^{d},
\end{equation}
To obtain a discrete Semantic-ID, we apply residual vector quantization with $L$ codebook levels. For level $\ell \in \{1,\dots,L\}$, we maintain a codebook:
\begin{equation}
\mathcal{C}^{(\ell)} = \{ c^{(\ell)}_1,\dots,c^{(\ell)}_{K_\ell} \}, \quad c^{(\ell)}_k \in \mathbb{R}^{d},
\end{equation}
At the first level, we choose the closest codeword to $z_i$:
\begin{equation}
q^{(1)}_i = \operatorname*{argmin}_{c \in \mathcal{C}^{(1)}} \|z_i - c\|_2^2,
\end{equation}
For higher levels, we iteratively quantize the residual:
\begin{equation}
r^{(\ell)}_i = z_i - \sum_{m=1}^{\ell-1} q^{(m)}_i,
\qquad
q^{(\ell)}_i = \operatorname*{argmin}_{c \in \mathcal{C}^{(\ell)}} \| r^{(\ell)}_i - c\|_2^2, \quad \ell = 2,\dots,L.
\end{equation}
The final quantized latent representation is the residual sum:
\begin{equation}
\hat{z}_i = \sum_{\ell=1}^{L} q^{(\ell)}_i.
\end{equation}
A decoder network $g_\phi(\cdot)$ reconstructs the input embedding:
\begin{equation}
\hat{x}_i = g_\phi(\hat{z}_i).
\end{equation}

\subsubsection{K-means Codebook Initialization}

To improve stability and avoid codebook collapse, each codebook $\mathcal{C}^{(\ell)}$ is initialized by k-means clustering. Specifically, we run k-means on the residuals at level $\ell$:
\begin{equation}
\mathcal{C}^{(\ell)} \leftarrow \operatorname{kmeans}(\{r^{(\ell)}_i\}_{i \in \mathcal{I}}, K_\ell),
\end{equation}
so that initial codewords reflect the global structure of item representations. After initialization, the codebooks are updated during training together with the encoder and decoder.

\subsubsection{Objective and Semantic-ID Tokens}

The RQ-VAE is trained with a reconstruction loss and a commitment loss:
\begin{equation}
\mathcal{L}_{\mathrm{RQ\text{-}VAE}}
=
\| x_i - \hat{x}_i \|_2^2
+
\beta \sum_{\ell=1}^{L}
\big\| \operatorname{sg}[z_i] - q^{(\ell)}_i \big\|_2^2,
\end{equation}
where $\operatorname{sg}[\cdot]$ denotes the stop-gradient operator, and $\beta$ controls the strength of the commitment term.

Each codeword index at level $\ell$ defines a discrete symbol in a shared vocabulary. For an item $i$, the sequence of selected codeword indices across levels:
\begin{equation}
\mathbf{c}_i = (c_{i,1}, \dots, c_{i,L_c}) \in \{1,\dots,V\}^{L_c},
\end{equation}
constitutes its \emph{Semantic-ID}, where $L_c$ is the Semantic-ID length and $V$ is the vocabulary size of discrete codes. These code sequences are used as tokens in the generative model.

\begin{table*}[!t]
\footnotesize
\setlength{\tabcolsep}{1pt}
\centering
\caption{Comparison of multi-step prediction performance (Seq-1, Seq-2, Seq-3) on the Beauty and Cell Phones \& Accessories datasets across all baseline methods and GCB. The best results are \underline{underlined} and $`-`$ indicates unavailable values.}
\vspace{-1em}
\label{tab:combined_results}
\begin{adjustbox}{max width=\textwidth}
\begin{tabular}{l|cccccc|cccccc|cccccc}
\toprule[1.5pt]
\textbf{Metrics} & \textbf{FPMC} & \textbf{SASRec} & \textbf{STOSA} & \textbf{PreferDiff} & \textbf{DCRec} & \textbf{GCB} & 
\textbf{FPMC} & \textbf{SASRec} & \textbf{STOSA} & \textbf{PreferDiff} & \textbf{DCRec} & \textbf{GCB} & 
\textbf{FPMC} & \textbf{SASRec} & \textbf{STOSA} & \textbf{PreferDiff} & \textbf{DCRec} & \textbf{GCB} \\
 & \multicolumn{6}{c|}{\textbf{Seq-1}} & \multicolumn{6}{c|}{\textbf{Seq-2}} & \multicolumn{6}{c}{\textbf{Seq-3}} \\
\midrule[1pt]
\multicolumn{19}{c}{\textbf{Beauty}} \\
\midrule[1pt]
MHR@5        & 0.0446 & 0.0327 & 0.0354 & 0.0049 & 0.0393 & \textbf{\underline{0.0454}} & 0.0255 & 0.0271 & 0.0219 & 0.0025 & \textbf{\underline{0.0300}} & 0.0276 & 0.0249 & 0.0211 & 0.0190 & 0.0012 & 0.0260 & \textbf{\underline{0.0259}} \\
MNDCG@5      & 0.0251 & 0.0267 & 0.0226 & 0.0025 & 0.0212 & \textbf{\underline{0.0321}} & 0.0138 & 0.0140 & 0.0115 & 0.0013 & 0.0160 & \textbf{\underline{0.0180}} & 0.0143 & 0.0114 & 0.0097 & 0.0008 & 0.0145 & \textbf{\underline{0.0154}} \\
MHR@10       & \textbf{\underline{0.0648}} & 0.0626 & 0.0615 & 0.0104 & 0.0605 & \textbf{\underline{0.0648}} & 0.0417 & 0.0413 & 0.0334 & 0.0043 & \textbf{\underline{0.0469}} & 0.0422 & 0.0386 & 0.0382 & 0.0288 & 0.0025 & 0.0514 & \textbf{\underline{0.0376}} \\
MNDCG@10     & 0.0272 & 0.0271 & 0.0227 & 0.0048 & 0.0243 & \textbf{\underline{0.0384}} & 0.0174 & 0.0164 & 0.0134 & 0.0019 & 0.0192 & \textbf{\underline{0.0227}} & 0.0159 & 0.0166 & 0.0115 & 0.0011 & \textbf{\underline{0.0248}} & 0.0191 \\
\midrule[1pt]
SHR@5        & 0.0446 & 0.0327 & 0.0449 & 0.0049 & 0.0393 & \textbf{\underline{0.0454}} & 0.0250 & 0.0268 & 0.0216 & 0.0022 & \textbf{\underline{0.0287}} & 0.0250 & 0.0233 & 0.0202 & 0.0186 & 0.0009 & 0.0241 & \textbf{\underline{0.0252}} \\
SNDCG@5      & 0.0251 & 0.0267 & 0.0226 & 0.0025 & 0.0212 & \textbf{\underline{0.0321}} & 0.0135 & 0.0139 & 0.0113 & 0.0011 & 0.0154 & \textbf{\underline{0.0167}} & 0.0131 & 0.0108 & 0.0094 & 0.0006 & 0.0135 & \textbf{\underline{0.0150}} \\
SHR@10       & \textbf{\underline{0.0648}} & 0.0626 & 0.0615 & 0.0104 & 0.0605 & \textbf{\underline{0.0648}} & 0.0408 & 0.0405 & 0.0330 & 0.0039 & \textbf{\underline{0.0454}} & 0.0374 & 0.0370 & 0.0359 & 0.0281 & 0.0017 & 0.0506 & \textbf{\underline{0.0358}} \\
SNDCG@10     & 0.0272 & 0.0271 & 0.0227 & 0.0048 & 0.0243 & \textbf{\underline{0.0384}} & 0.0171 & 0.0160 & 0.0133 & 0.0017 & 0.0187 & \textbf{\underline{0.0207}} & 0.0152 & 0.0151 & 0.0111 & 0.0007 & \textbf{\underline{0.0240}} & 0.0183 \\
\midrule[1pt]
1st\_HR@5    & 0.0446 & 0.0327 & 0.0449 & 0.0049 & 0.0393 & \textbf{\underline{0.0454}} & 0.0304 & 0.0313 & 0.0256 & 0.0037 & 0.0388 & \textbf{\underline{0.0392}} & 0.0260 & 0.0277 & 0.0238 & 0.0009 & \textbf{\underline{0.0389}} & 0.0333 \\
1st\_NDCG@5  & 0.0251 & 0.0267 & 0.0226 & 0.0025 & 0.0212 & \textbf{\underline{0.0321}} & 0.0165 & 0.0161 & 0.0137 & 0.0019 & 0.0204 & \textbf{\underline{0.0247}} & 0.0146 & 0.0151 & 0.0125 & 0.0005 & \textbf{\underline{0.0216}} & 0.0208 \\
1st\_HR@10   & \textbf{\underline{0.0648}} & 0.0626 & 0.0615 & 0.0104 & 0.0605 & \textbf{\underline{0.0648}} & 0.0503 & 0.0494 & 0.0384 & 0.0059 & 0.0588 & \textbf{\underline{0.0618}} & 0.0415 & 0.0422 & 0.0374 & 0.0033 & \textbf{\underline{0.0625}} & 0.0549 \\
1st\_NDCG@10 & 0.0272 & 0.0271 & 0.0227 & 0.0048 & 0.0243 & \textbf{\underline{0.0384}} & 0.0210 & 0.0199 & 0.0152 & 0.0025 & 0.0236 & \textbf{\underline{0.0319}} & 0.0168 & 0.0174 & 0.0148 & 0.0014 & 0.0264 & \textbf{\underline{0.0277}} \\
\midrule[1pt]
2nd\_HR@5    & ---    & ---    & ---    & ---    & ---    & ---    & 0.0206 & \textbf{\underline{0.0229}} & 0.0182 & 0.0014 & 0.0213 & 0.0160 & 0.0348 & 0.0223 & 0.0186 & 0.0024 & 0.0244 & \textbf{\underline{0.0183}} \\
2nd\_NDCG@5  & ---    & ---    & ---    & ---    & ---    & ---    & 0.0110 & \textbf{\underline{0.0120}} & 0.0094 & 0.0007 & 0.0116 & 0.0112 & \textbf{\underline{0.0207}} & 0.0123 & 0.0097 & 0.0014 & 0.0136 & 0.0118 \\
2nd\_HR@10   & ---    & ---    & ---    & ---    & ---    & ---    & 0.0331 & \textbf{\underline{0.0332}} & 0.0284 & 0.0027 & 0.0351 & 0.0227 & 0.0501 & 0.0505 & 0.0281 & 0.0039 & 0.0407 & \textbf{\underline{0.0278}} \\
2nd\_NDCG@10 & ---    & ---    & ---    & ---    & ---    & ---    & \textbf{\underline{0.0138}} & 0.0129 & 0.0116 & 0.0012 & 0.0148 & 0.0134 & \textbf{\underline{0.0242}} & 0.0207 & 0.0117 & 0.0016 & 0.0168 & 0.0148 \\
\midrule[1pt]
3rd\_HR@5    & ---    & ---    & ---    & ---    & ---    & ---    & ---    & ---    & ---    & ---    & ---    & ---    & 0.0139 & 0.0134 & 0.0145 & 0.0004 & 0.0147 & \textbf{\underline{0.0262}} \\
3rd\_NDCG@5  & ---    & ---    & ---    & ---    & ---    & ---    & ---    & ---    & ---    & ---    & ---    & ---    & 0.0075 & 0.0067 & 0.0069 & 0.0004 & 0.0084 & \textbf{\underline{0.0136}} \\
3rd\_HR@10   & ---    & ---    & ---    & ---    & ---    & ---    & ---    & ---    & ---    & ---    & ---    & ---    & 0.0244 & 0.0218 & 0.0210 & 0.0004 & \textbf{\underline{0.0580}} & 0.0300 \\
3rd\_NDCG@10 & ---    & ---    & ---    & ---    & ---    & ---    & ---    & ---    & ---    & ---    & ---    & ---    & 0.0102 & 0.0082 & 0.0079 & 0.0001 & \textbf{\underline{0.0311}} & 0.0148 \\
\midrule[1pt]
\multicolumn{19}{c}{\textbf{Cell Phones \& Accessories}} \\
\midrule[1pt]
MHR@5        & 0.0385 & 0.0535 & 0.0423 & 0.0092 & 0.0413 & \textbf{\underline{0.0637}} & 0.0171 & 0.0163 & 0.0123 & 0.0048 & 0.0086 & \textbf{\underline{0.0261}} & 0.0128 & 0.0147 & 0.0122 & 0.0034 & 0.0185 & \textbf{\underline{0.0252}} \\
MNDCG@5      & 0.0230 & 0.0283 & 0.0220 & 0.0049 & 0.0227 & \textbf{\underline{0.0408}} & 0.0099 & 0.0091 & 0.0067 & 0.0027 & 0.0052 & \textbf{\underline{0.0167}} & 0.0075 & 0.0080 & 0.0067 & 0.0017 & 0.0098 & \textbf{\underline{0.0156}} \\
MHR@10       & 0.0637 & 0.0782 & 0.0596 & 0.0167 & 0.0672 & \textbf{\underline{0.0965}} & 0.0306 & 0.0240 & 0.0189 & 0.0089 & 0.0177 & \textbf{\underline{0.0411}} & 0.0220 & 0.0219 & 0.0175 & 0.0060 & 0.0317 & \textbf{\underline{0.0353}} \\
MNDCG@10     & 0.0264 & 0.0306 & 0.0226 & 0.0078 & 0.0279 & \textbf{\underline{0.0514}} & 0.0133 & 0.0094 & 0.0074 & 0.0039 & 0.0080 & \textbf{\underline{0.0215}} & 0.0098 & 0.0091 & 0.0071 & 0.0024 & 0.0135 & \textbf{\underline{0.0188}} \\
\midrule[1pt]
SHR@5        & 0.0385 & 0.0535 & 0.0423 & 0.0092 & 0.0413 & \textbf{\underline{0.0637}} & 0.0169 & 0.0162 & 0.0123 & 0.0047 & 0.0086 & \textbf{\underline{0.0239}} & 0.0126 & 0.0144 & 0.0119 & 0.0033 & 0.0176 & \textbf{\underline{0.0231}} \\
SNDCG@5      & 0.0230 & 0.0283 & 0.0220 & 0.0049 & 0.0227 & \textbf{\underline{0.0408}} & 0.0098 & 0.0091 & 0.0067 & 0.0027 & 0.0052 & \textbf{\underline{0.0154}} & 0.0074 & 0.0078 & 0.0065 & 0.0017 & 0.0092 & \textbf{\underline{0.0145}} \\
SHR@10       & 0.0637 & 0.0782 & 0.0596 & 0.0167 & 0.0672 & \textbf{\underline{0.0965}} & 0.0303 & 0.0239 & 0.0188 & 0.0087 & 0.0176 & \textbf{\underline{0.0370}} & 0.0218 & 0.0211 & 0.0168 & 0.0059 & 0.0305 & \textbf{\underline{0.0322}} \\
SNDCG@10     & 0.0264 & 0.0306 & 0.0226 & 0.0078 & 0.0279 & \textbf{\underline{0.0514}} & 0.0131 & 0.0094 & 0.0074 & 0.0039 & 0.0080 & \textbf{\underline{0.0196}} & 0.0097 & 0.0088 & 0.0068 & 0.0024 & 0.0129 & \textbf{\underline{0.0174}} \\
\midrule[1pt]
1st\_HR@5    & 0.0385 & 0.0535 & 0.0423 & 0.0092 & 0.0413 & \textbf{\underline{0.0637}} & 0.0194 & 0.0172 & 0.0127 & 0.0054 & 0.0089 & \textbf{\underline{0.0364}} & 0.0132 & 0.0164 & 0.0141 & 0.0026 & 0.0255 & \textbf{\underline{0.0388}} \\
1st\_NDCG@5  & 0.0230 & 0.0283 & 0.0220 & 0.0049 & 0.0227 & \textbf{\underline{0.0408}} & 0.0110 & 0.0097 & 0.0070 & 0.0030 & 0.0054 & \textbf{\underline{0.0229}} & 0.0077 & 0.0082 & 0.0078 & 0.0014 & 0.0142 & \textbf{\underline{0.0224}} \\
1st\_HR@10   & 0.0637 & 0.0782 & 0.0596 & 0.0167 & 0.0672 & \textbf{\underline{0.0965}} & 0.0355 & 0.0257 & 0.0205 & 0.0106 & 0.0197 & \textbf{\underline{0.0589}} & 0.0241 & 0.0270 & 0.0201 & 0.0070 & 0.0441 & \textbf{\underline{0.0547}} \\
1st\_NDCG@10 & 0.0264 & 0.0306 & 0.0226 & 0.0078 & 0.0279 & \textbf{\underline{0.0514}} & 0.0154 & 0.0101 & 0.0081 & 0.0048 & 0.0091 & \textbf{\underline{0.0302}} & 0.0103 & 0.0115 & 0.0080 & 0.0033 & 0.0192 & \textbf{\underline{0.0275}} \\
\midrule[1pt]
2nd\_HR@5    & ---    & ---    & ---    & ---    & ---    & ---    & 0.0148 & 0.0153 & 0.0119 & 0.0041 & 0.0083 & \textbf{\underline{0.0157}} & 0.0156 & 0.0174 & 0.0138 & 0.0039 & 0.0184 & \textbf{\underline{0.0234}} \\
2nd\_NDCG@5  & ---    & ---    & ---    & ---    & ---    & ---    & 0.0088 & 0.0085 & 0.0064 & 0.0024 & 0.0050 & \textbf{\underline{0.0104}} & 0.0093 & 0.0102 & 0.0074 & 0.0021 & 0.0093 & \textbf{\underline{0.0158}} \\
2nd\_HR@10   & ---    & ---    & ---    & ---    & ---    & ---    & \textbf{\underline{0.0258}} & 0.0223 & 0.0173 & 0.0071 & 0.0156 & 0.0233 & 0.0235 & 0.0247 & 0.0213 & 0.0057 & 0.0289 & \textbf{\underline{0.0319}} \\
2nd\_NDCG@10 & ---    & ---    & ---    & ---    & ---    & ---    & 0.0112 & 0.0087 & 0.0067 & 0.0031 & 0.0070 & \textbf{\underline{0.0128}} & 0.0103 & 0.0100 & 0.0089 & 0.0022 & 0.0120 & \textbf{\underline{0.0185}} \\
\midrule[1pt]
3rd\_HR@5    & ---    & ---    & ---    & ---    & ---    & ---    & ---    & ---    & ---    & ---    & ---    & ---    & 0.0097 & 0.0105 & 0.0088 & 0.0037 & 0.0116 & \textbf{\underline{0.0135}} \\
3rd\_NDCG@5  & ---    & ---    & ---    & ---    & ---    & ---    & ---    & ---    & ---    & ---    & ---    & ---    & 0.0056 & 0.0057 & 0.0049 & 0.0016 & 0.0058 & \textbf{\underline{0.0086}} \\
3rd\_HR@10   & ---    & ---    & ---    & ---    & ---    & ---    & ---    & ---    & ---    & ---    & ---    & ---    & 0.0184 & 0.0141 & 0.0111 & 0.0051 & \textbf{\underline{0.0222}} & 0.0192 \\
3rd\_NDCG@10 & ---    & ---    & ---    & ---    & ---    & ---    & ---    & ---    & ---    & ---    & ---    & ---    & 0.0088 & 0.0060 & 0.0044 & 0.0018 & 0.0095 & \textbf{\underline{0.0104}} \\
\bottomrule[1.5pt]
\end{tabular}
\end{adjustbox}
\end{table*}

\subsection{Generative User Behaviour Trajectory Prediction}
\subsubsection{Tokenization of User Histories and Futures}
Given a user $u$ with interaction history:
\begin{equation}
H_u = (i_{u,1}, i_{u,2}, \dots, i_{u,n}),
\end{equation}

we convert each item $i_{u,j}$ into its Semantic-ID sequence $\mathbf{c}_{i_{u,j}} \in \{1,\dots,V\}^{L_c}$. Concatenating all Semantic-IDs yields the tokenized history:
\begin{equation}
X_u = (\mathbf{c}_{i_{u,1}}, \dots, \mathbf{c}_{i_{u,n}}) \in \{1,\dots,V\}^{L_h},
\end{equation}
where $L_h = n \cdot L_c$ is the history token length (after padding and truncation to a maximum length if necessary). Similarly, the future sequence:
\begin{equation}
S_u = (i_{u,n+1}, \dots, i_{u,n+k}),
\end{equation}
is mapped to its tokenized target sequence:
\begin{equation}
Y_u = (\mathbf{c}_{i_{u,n+1}}, \dots, \mathbf{c}_{i_{u,n+k}}) \in \{1,\dots,V\}^{L_y},
\end{equation}
with $L_y = k \cdot L_c$. In practice, we append a special end-of-sequence token and use padding tokens so that all sequences share a fixed maximum length.

We instantiate the generative model as a T5-style encoder–decoder architecture \cite{rajput2023recommender}, denoted as $\mathrm{GCB}$. The encoder takes the tokenized history $X_u$ and its attention mask as input and produces contextualized representations. The decoder autoregressively generates the target token sequence $Y_u$ conditioned on both the encoder outputs and previously generated tokens.

Formally, GCB models the conditional distribution:
\begin{equation}
p_\theta(Y_u \mid X_u) = \prod_{t=1}^{L_y}
p_\theta\big( y_{u,t} \mid y_{u,<t}, X_u \big),
\end{equation}
where $y_{u,t}$ is the $t$-th target token and $y_{u,<t}$ denotes all previous target tokens. During training, we use teacher forcing and maximize the log-likelihood of the ground-truth token sequence. The training objective over a dataset $\mathcal{D}$ of users is:
\begin{equation}
\mathcal{L}_{\mathrm{GCB}}
=
- \frac{1}{|\mathcal{D}|}
\sum_{u \in \mathcal{D}}
\sum_{t=1}^{L_y}
\log p_\theta\big( y_{u,t} \mid y_{u,<t}, X_u \big),
\end{equation}
The model is optimized using Adam with a fixed learning rate, and early stopping is applied based on validation performance.

\subsubsection{Beam Search Generation in Semantic-ID Space}

At inference time, GCB generates future Semantic-ID sequences using beam search. Given an input history $X_u$, the model starts from the decoder start token and expands candidate token sequences by repeatedly selecting the most likely next token. With beam size $B$, this procedure yields $B$ candidate token sequences:
\begin{equation}
\mathcal{Y}_u = \{\hat{Y}_u^{(1)}, \dots, \hat{Y}_u^{(B)}\},
\end{equation}
where each $\hat{Y}_u^{(b)} \in \{1,\dots,V\}^{\hat{L}_y}$ is a generated token sequence of length up to a maximum $\hat{L}_y$.

We partition each generated token sequence $\hat{Y}_u^{(b)}$ into consecutive blocks of length $L_c$, each block representing a predicted Semantic-ID for one future item. The first $k$ blocks define a candidate future trajectory:
\begin{equation}
\hat{S}_u^{(b)} = (\hat{i}_{u,n+1}^{(b)}, \dots, \hat{i}_{u,n+k}^{(b)}).
\end{equation}
where each $\hat{i}_{u,n+j}^{(b)}$ is obtained by mapping the corresponding Semantic-ID block back to an item index if a valid mapping exists.

To ensure validity, we restrict attention to beams whose Semantic-ID sequences can be mapped back to items via a precomputed dictionary ($\texttt{code\_to\_item}$). Invalid beams that cannot be decoded into valid items are discarded before metric computation.


\section{Experiments}


\textbf{Datasets.} We evaluate GCB on two Amazon review datasets \citep{AmazonDataset}: Beauty and Cell Phones \& Accessories. We filter out users with fewer than $3 \times \text{TARGET\_LEN} + 1$ interactions so that each user has sufficient history for multi-step prediction and for constructing training, validation, and test segments. We adopt a leave-$k$-out splitting strategy with $k=\text{TARGET\_LEN}$. For a user sequence $H_u$, we form three prediction segments: (1) the training input is the prefix $H_u[1:|H_u|-2k]$ with the next $k$ items as targets; (2) the validation input is $H_u[1:|H_u|-k]$ with the subsequent $k$ items as targets; and (3) the test input uses all but the last $k$ items to predict the final $k$ interactions.
We conduct experiments with three different target lengths: $\text{TARGET\_LEN} \in \{1, 2, 3\}$, corresponding to predicting 1, 2, or 3 future items, respectively. In Table~\ref{tab:combined_results}, results are organized into three groups (\textbf{Seq-1}, \textbf{Seq-2}, \textbf{Seq-3}) based on the target length. For each setting, we report aggregated metrics (MHR, MNDCG for mean across positions; SHR, SNDCG for sequence-level evaluation) as well as position-specific metrics (e.g., 1st\_HR@5, 2nd\_HR@5, 3rd\_HR@5) to measure accuracy at each individual future step.

\textbf{Baselines.} We conduct a comprehensive comparison of GCB against five representative baselines from sequential modeling. These include three conventional methods: FPMC \cite{FPMC}, SASRec \cite{SARS}, STOSA \cite{fan2022sequential}, two SOTA
diffusion-based generative methods: PreDiff \cite{liu2024preference}, DCRec \cite{huang2024dual}.

\textbf{Evaluation Metrics.} To assess the entire predicted trajectory, we adapt the standard next-item metrics HR@K and NDCG@K into sequence-level variants: SeqHR and SeqNDCG. These metrics evaluate the sequence as a whole by taking the geometric mean of the hit rates or NDCG scores computed at each prediction step.

\textbf{Implementation Details.} In our setting, GCB employs a hierarchical semantic ID space with code length $4$. The first three code positions are learned through RQ-VAE followed by k-means refinement on latent embeddings, while the fourth position is reserved for collision avoidance. The autoregressive generator is a customized T5-based transformer with $4$ encoder layers and $4$ decoder layers, hidden size $128$, feed-forward dimension $1024$, and $6$ attention heads. We use Adam with a learning rate of $\mathbf{0.0001}$ and batch size $256$. Training runs for up to $300$ epochs with early stopping based on validation metrics, and a warm-up phase of $50$ epochs without evaluation to stabilize code learning. For inference, we apply beam search with a beam size of $100$ to generate multi-step trajectories. All experiments are conducted on NVIDIA L40S GPUs, and reported results are averaged over multiple runs for robustness.

\begin{figure}[h]
    \centering
    \begin{subfigure}{1\linewidth}
        \centering
        \includegraphics[width=\linewidth]{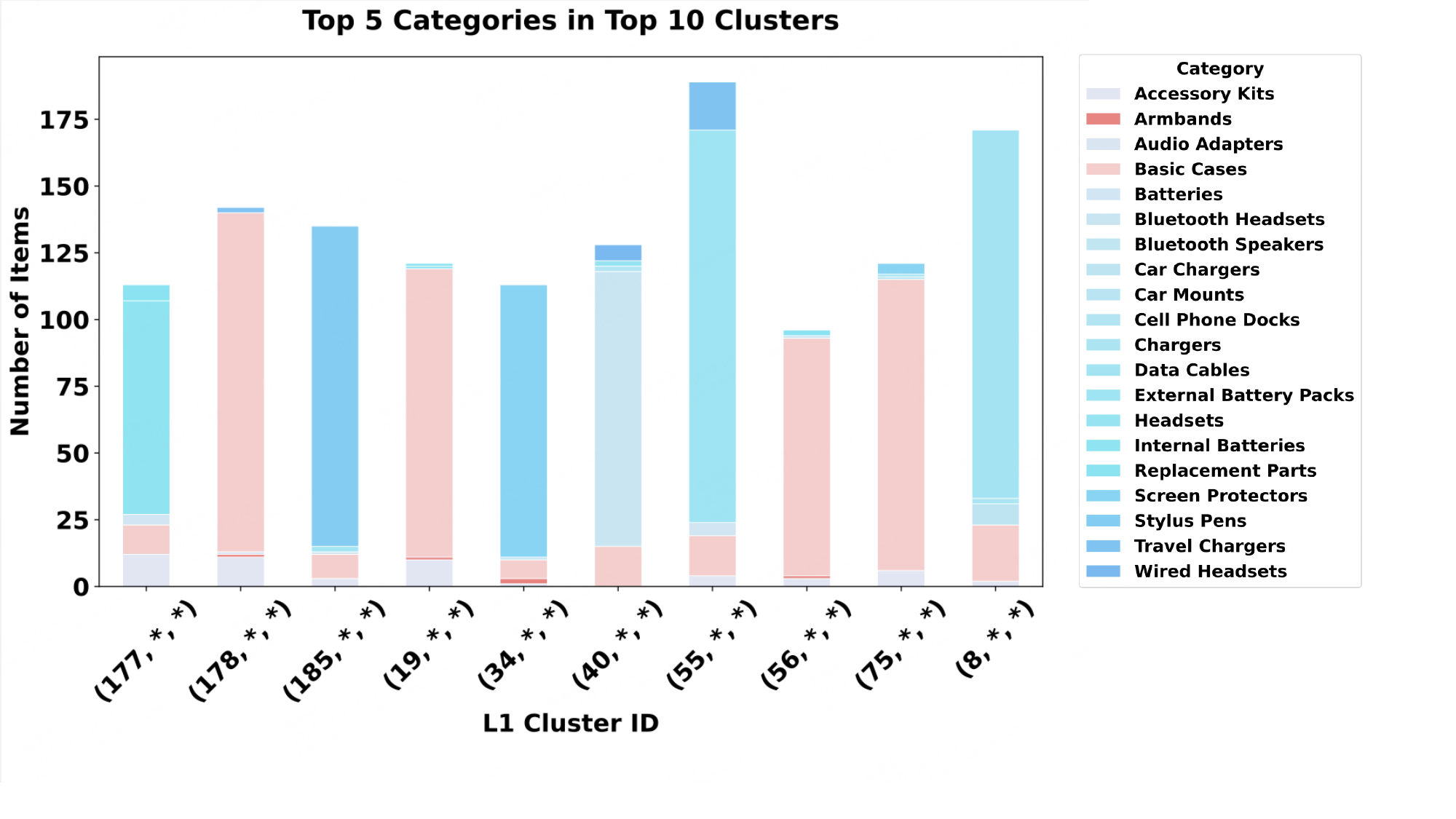}
        \vspace{-2em}
        \caption{Category distribution for all items in L1 clusters.}
        \label{fig:img1}
    \end{subfigure}
        \vspace{0.5cm} 
    \begin{subfigure}{1\linewidth}
        \centering
        \includegraphics[width=\linewidth]{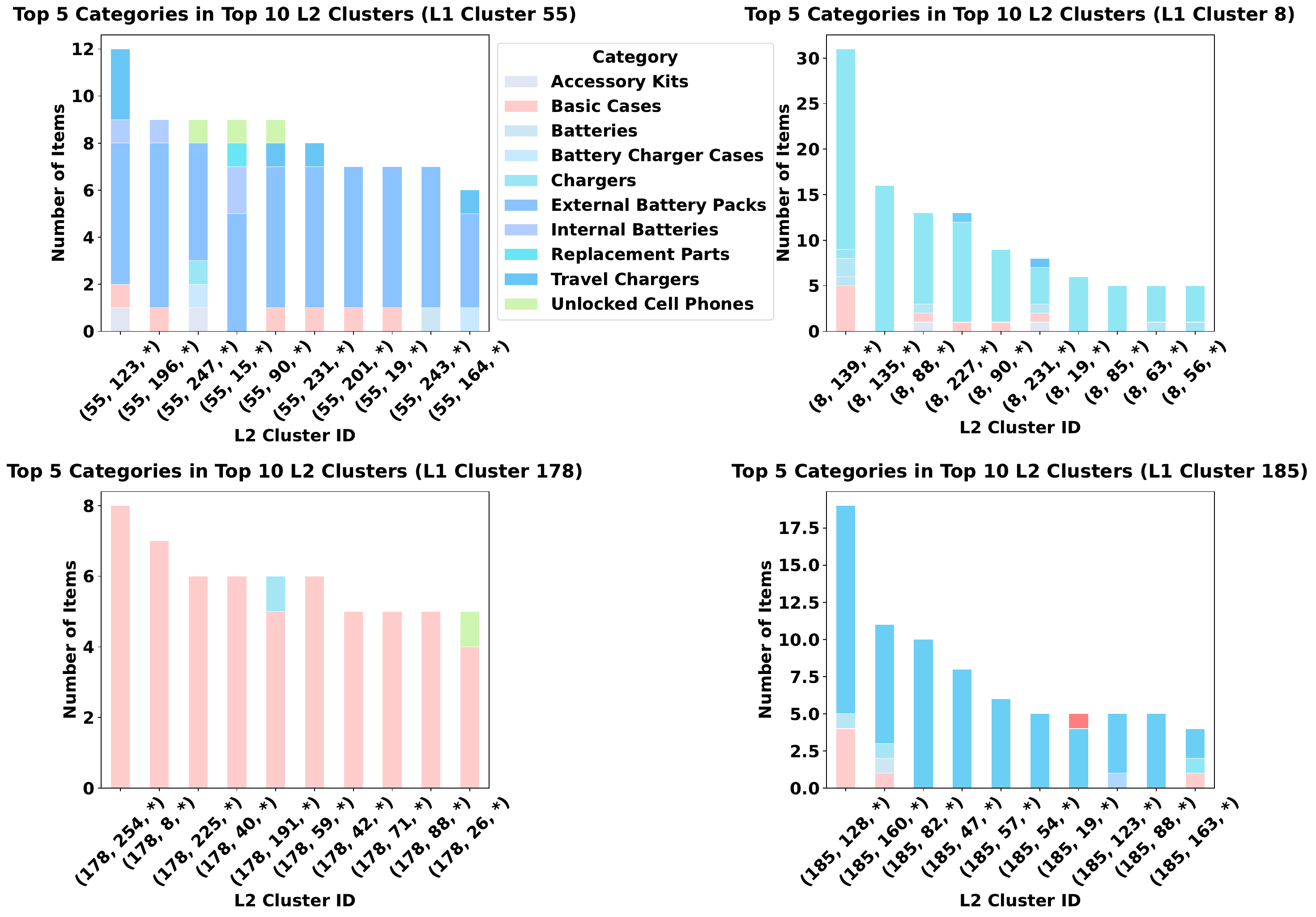}
        \caption{Category distribution for top L2 clusters within each L1 cluster.}
        \label{fig:img2}
    \end{subfigure}
    \vspace{-3em}
    \caption{Overall category distribution in L1 and L2 clusters.}
    \vspace{-1em}
    \label{fig:two_images}
\end{figure}

\textbf{Main Results.} 
Across both the Beauty and Cell Phones datasets, our GCB consistently achieves the best or near-best performance on the majority of multi-step top-$K$ metrics. In the first prediction step (Seq-1), GCB clearly surpasses classical sequential models (FPMC, SASRec), attention-based STOSA, and recent generative baselines (PreferDiff, DCRec), showing notably higher HitRate and NDCG. As the prediction horizon becomes longer (Seq-2, Seq-3), all models experience performance degradation, yet GCB remains highly competitive and often maintains the leading results, especially on Cellphones, where the gains are most pronounced. These findings demonstrate that representing items through Semantic-ID codes and generating future trajectories in this compact space substantially enhances both ranking accuracy and robustness in multi-step recommendation.

\textbf{Hierarchical Semantic Structure Analysis.} Figure~\ref{fig:two_images} illustrates the semantic structure learned by our RQ-Kmeans encoding. At the L1 level, clusters naturally group items into coherent high-level categories (e.g., cases, chargers, audio accessories), showing that the first-stage codes capture broad functional similarity. Within each L1 cluster, the L2 sub-clusters further refine this structure into more specific item types, such as different case styles or charger variants. This hierarchical organization reveals that our semantic IDs preserve meaningful item relationships, offering both interpretability and a well-structured discrete space that facilitates more accurate generative trajectory prediction.

\section{Conclusion}

In this paper, we introduced Generative Chain of Behavior (GCB), a unified generative framework for modeling long-term user behavior trajectories. By constructing a semantic ID space with RQ-VAE and k-means refinement, GCB maps items into a structured discrete representation that preserves semantic proximity. Building on this space, a transformer-based autoregressive generator produces multi-step future interactions as coherent behavioral chains, capturing complex intent transitions beyond traditional next-item prediction. Experiments on multiple benchmarks show that GCB achieves state-of-the-art performance in multi-step forecasting while enhancing the semantic precision and consistency of generated trajectories.


\bibliographystyle{ACM-Reference-Format}
\bibliography{sample-base}

\end{document}